# Observation of the Kibble–Zurek Mechanism in Microscopic Acoustic Crackling Noises

H.O. Ghaffari [1(A)], P. Benson[2], W. A. Griffth[1], K. Xia[3], R.P. Young[3]

[1] The University of Texas at Arlington, 500 Yates St. Arlington, TX 76019

[2] Rock Mechanics Laboratory, School of Earth and Environmental Sciences, University of Portsmouth, Burnaby building, Portsmouth, PO1 3QL, UK

[3] Department of Civil Engineering and Lassonde Institute, University of Toronto, Toronto, 170 College Street, M5s3e3, On, Canada

**Characterizing the fast evolution of microstructure is key to understanding "crackling" phenomena during the deformation of solid brittle materials. For example, it has been proposed using atomistic simulations of crack propagation in elastic materials that the formation of a nonlinear hyperelastic zone around moving crack tips controls crack velocity. To date, progress in understanding the physics of this critical zone has been limited due to the lack of data describing the complex physical processes that operate near microscopic crack tips. We show, by analyzing acoustic emissions during rock deformation experiments, that the signature of this nonlinear zone maps directly to crackling noises. In particular, we use a functional network method to characterize a weakening zone that forms near the moving crack tips, and we determine the scaling law between the formation of defects and the traversal rate across the critical point of transition to rapid weakening at the crack tip. Our results are in agreement with the Kibble-Zurek Mechanism (KZM).**

The spatio-temporal evolution of acoustic signals, known as crackling noise [1], is a direct result of failing atomic bonds during material fracture. Such signals, if properly interpreted, may be used to better understand the dynamics of rupture progress in the vicinity of crack tips over a broad range of scales and conditions [2-4]. During crackling, part of the stored energy near the crack tip is consumed in the breaking of molecular and atomic bonds, resulting in new crack surface(s). The key to understanding the crackling process lies in the characterizing structure of the near-tip region, where stresses amplitudes are large. Due to the microscopic size and high speeds encountered in the area of the crack tip, direct measurements are difficult, and analysis typically relies on computational techniques [5]. Because of the large strains present near crack tips, nonlinear elastic contributions must occur. Recent work [2-3] suggests that non-linearity around the moving crack tip governs the rupture velocity. Specifically, the local hyperelastic zone around the moving crack tip enhances energy flow for stiffening systems, and reduces energy flow for softening systems, resulting respectively in increases and decreases in the fracture velocity relative to the linear elastic case [2]. Furthermore, investigations using "*slow*" cracks in gels demonstrate a link between the spatial energy flow around the rupture tip and the curvature of the tip [5]. This link is thought to be responsible for inaccuracies in linear elastic analyses that are commonly used in material science for simulating crack tip processes, with applications ranging from metal fatigue to earthquake nucleation.

In this study we use a functional network method [6-7] applied to acoustic emission (AE) data recorded from rock deformation and rock friction experiments (see methods section) to show that moving micro-cracks contain signatures of non-linearity. We discover that the onset of the non-linear stage prior to global failure coincides with the nucleation of kink instabilities, and we use "*network strings*" to visualize spatio-temporal evolution of these topological defects. For the first time, we show that emitted crackling noises hold the signature of the Kibble-Zurek mechanism (KZM) [8-11] that provides an estimation of the defect density as a function of the traversing rate. A key output of KZM relates faster ramp rates to higher defect density, as the result of a spontaneous symmetry breaking process. We measure real-time evolution of a first-order correlation function of the system (networks) and verify the main prediction of KZM, namely the power-law scaling of the correlation length with the ramp rate. Moreover, we show that the correlation length near the transition remain effectively frozen. This is the main underlying hypothesis behind the Kibble-Zurek mechanism. In addition, using our laboratory data sets we illustrate that adiabatic-impulse transition, as the core of the KZ hypothesis, plays a major role in approximating weakening rate (i.e., rupture velocity[4]).

We analyse AE waveforms (the laboratory analogue of seismograms due to rock fracture or earthquake rupture) under different simulated depth (pressure) conditions and loading paths, and on two rock types: Westerly granite and Basalt from Mount Etna, Italy. (datasets Lab.EQ1, Lab.EQ2, Lab.EQ3 and Lab.EQ4 - see methods section and supporting information). We apply tools from the theory of complex networks to analyze emitted noises from microscopic cracks, where the acoustic time series recorded at each sensor is represented as a node [SI-section 1; Ref.6,7]. These results allow us to develop an interpretation of reordered multiple acoustic-crackling signals involving a microsecond evolution of different dynamic crack tip phases as encoded in the network modularity ($Q$) which we refer to as "$Q$-profiles" (see methods section). This evolution can be broken down into three distinct phases (Fig. 1a, Fig. 1d-e): (1). The S-phase: an initial strengthening phase preceding the critical point at which point weakening and catastrophic failure begins; (2) the W-phase: a fast-slip or weakening phase; and (3) the D-phase: a slow slip or decelerating phase [6-7,12]. To better understand the S-phase and how transition occurs over the critical point between S to W, we use the reciprocal of modularity ($R = Q^{-1}$) profiles (i.e., "$R$-profiles") which closely resemble dynamic stress profiles commonly used to characterize rock failure (Fig. 1). In the following discussion, we describe the observation of defect formation prior to onset of the W phase. To proceed, we define the critical zone onset, $R_c$, as the value of R at the time of the first impulse in the inverse of mean betweenness centrality ($1/\log\langle B.C.\rangle$) profile, where $\langle\ \rangle$ indicates the mean value of all nodes (Fig.2 & 3). In Fig.2, we show $1/\log\langle B.C.\rangle$ profiles for 6 acoustic events from our laboratory tests. For all events this profile is characterized by a narrow impulse at the transition from the S to W phase.

Later, we will show that the first impulse corresponds to the first nucleated defect in transition from S to W while the second spike indicates the defects' formation in an inverse transition (W to S). We define the duration, $\ddagger_c$, as the temporal length of this impulse zone in transition from S to W. This also corresponds to the time between $R_c$ and $R_{max}$, the maximum value of R(t). $R_{max}$ is the critical point, where the failure occurs and fast-weakening onsets. The duration of defect formation (nucleation time) $\ddagger_c$ varies between 0.25 µs to 8 µs.

In order to study the spatial variability of this impulse regime, we visualize the spatial evolution of the degree $k_i$ of the $i^{th}$ node (see supplementary material and [4]), (in 2D) using polar coordinates $(r_i, \theta_i)_{i=1,...,Nodes}$ where $r_i = k_i$ and $\theta_i$ indicates the position of the node (which is fixed on the outer circumference of the cylindrical sample; Figure 3c). We refer to these configurations as "$K$-strings", and the normal vector of the $K$-strings at each node indicates the local direction of increasing $k_i$ with time. We evaluate the variation of $r_i = k_i$ at each position (node) while we consider the temporal evolution of each single event (Fig.3; Fig.S5-S6). In Fig 3, we show that the onset of the impulse zone coincides with the folding of $K$-strings where the normal vectors are flipped at the onset of the non-linear regime and form a local domain that we refer to as a "kink" (Fig. 3d). Therefore, formation of domains in the course of the S-W transition results in a non-linear behaviour in the *S* phase of the *R*-profile. Since the trend of R(*t*) does match with the mean value of *k* for all nodes $\langle k(t) \rangle$ (Fig.S4): $\langle k(t) \rangle \sim R(t) \sim \dagger(t)$, then we infer $\langle k(t) \rangle \sim \dagger(t)$. Then, the non-linearity of stress is well-connected to nucleation of these kinks prior to onset of the W-phase where fast-weakening occurs. The W-phase corresponds to global instability of K-strings and "collapsing" of K-string occurs where almost all nodes $\frac{\partial k_i}{\partial t} < 0$ (in other words, where all normal vectors of the K-strings point inward).

As emphasized by Polyakov [13] in his string theory, crumpled stings are analogous with the Heisenberg paramagnet while undulations destroy long-range order in surface normals [13-14]. These defects are local defects in initially ordered structures and can be removed by global collapse of K-strings and local bending or twisting around the defects cannot remove them (i.e., topological defects)[15,16]. To analysis deformation of K-strings, we map them on the "Ising chain" where for each node we assign $s_i = sign(\frac{\partial k_i}{\partial t})$ and then $s_i = \pm 1$. With this mapping, defects are represented with negative $s_i$ indicating flipped, inward-pointing normal vectors. A kink separating zones with up and down nodes (in analogy with "spins") functions as a locator for the change from one ground state (S-phase) to another degenerate ground state (W-phase) [17]. Then, we can study the real-time evolution of a cracking noise with a typical Ising chain when the K-string goes through a phase transition.

The critical point $\}_c$ is defined when $m = \langle s \rangle \to 0$ which coincides with $R_{max}$ (Fig.4, 5b). Here, $m$ is the *order parameter* of the K-strings and transition from S to W (and vice versa) occurs continuously. Furthermore, we have calculated a correlation function for the K-strings including all nodes approaching the critical point (Fig.4 a,b). We can fit a correlation function such as $G(x) = (1 - \frac{x}{L}) \exp(-\frac{x}{‹})$ where $L$ is the total number of nodes, $x$ is distance and ‹ is the correlation length. Correlation length ‹ is the cut-off length of the correlation function where for shorter distances than the correlation length, a power law function can be fitted. For a fully ordered state a triangular function (i.e. fully coherent system and ‹ $\to \infty$) is given by green line in Figure 4c.

Adding defects destroys long range of order among nodes (Fig.4b-Fig.S14-15). Interestingly, when approaching $m \approx 0$, the correlation length becomes essentially frozen as shown in Fig.4d for crackling event #255. This means as the system approaches the critical point at a finite rate, and after some point in time the correlation length cannot follow its diverging equilibrium value and transition occurs with $\xi$ much smaller than the system size. This correlation length sets the mean finite-size of final domains. The formation of domains as well as observation of frozen correlation length at the critical point of our acoustic networks indicate the signature of a spontaneous symmetry breaking process, *i.e.,* Kibble-Zurek mechanism (KZM, see Methods section) [8-10]. The core idea behind the Kibble-Zurek mechanism is that near the transition, the freezing of the correlation length is unavoidable. Based on this theory, the resulting density of defects left behind by continuous transitions is dependent on the rate at which the critical point is traversed, and the rate with which the system can adjust (the relaxation time). This mechanism is reflected in the density of defects and "*freeze-out*" time which scales with ramp-rate. To test the KZM's density of defect prediction, we carefully measured the number of flipped nodes in the vicinity of the critical point where correlation length is frozen (Fig. 6).

A key output of our analysis is that the number of defects (i.e., flipped nodes at the final state) we observe is larger when the local ramp rate $\frac{dR}{dt}$ is faster (Fig. 6a). We can fit a power-law scaling as: $\hat{\xi} \sim (\frac{dR}{dt})_{t=t_c}^{-0.35}$ (Fig.6a). Here $\hat{\xi}$ is the frozen correlation length and $(\frac{dR}{dt})_{t=t_c}$ is the ramp rate (Fig.S4-6,S10). To measure the ramp rate, we measure the slope of $R(t)$ prior to the impulse zone where we have fully a coherent system (Fig. 3b); this will yield the ramp rate, which is analogous to the local loading rate prior to the nucleation of kinks. The exponent of ∼*0.35 ±0.06* obtained by fitting the laboratory data is in agreement (within experimental error) with the theoretical value $\epsilon/(1+\epsilon z) \cong 0.34$ where, for a mean field model for 1D-kinks, $\epsilon \cong 0.5$ and $z \cong 1$ [18]. Here, the parameters $\epsilon$ and $z$ are spatial and dynamical critical exponents (see Methods section).

Furthermore, we can evaluate the time-reversal transition from the W to S phase while we approach to the critical point from right (in other words, if we heal or reverse the failure process). Approaching from the left (S-phase) or right (W-phase; time reversal or healing scenario) to the critical point $\tau_c$ results in slightly different characteristics of the defects' density (Fig.5b). The rate of the S-ramp (linear strengthening rate or ramp rate) for most of the recorded events is higher than the rate of W-ramp (i.e., linear weakening rate). Next we estimate the linear weakening rate. Using the KZM scaling law for defects, we obtain (Supplementary Information): $\tau_w \sim \tau_0 [1-(\frac{\tau_s}{\tau_0})^{\frac{-2\epsilon}{1+\epsilon z}}]^{\frac{1+\epsilon z}{-2\epsilon}}$, in which $\tau_0$ is determined by the microscopic details of the system; $\tau_s$ and $\tau_w$ are the time characteristics of ramps in the S and W phases, respectively. Therefore, the linear weakening rate is given by: $(\frac{dR}{dt})_w \sim \tau_0 [1 - (\frac{\tau_s}{\tau_0})^{\frac{-2\epsilon}{1+\epsilon z}}]^{\frac{1+\epsilon z}{2\epsilon}}$. A faster local S-ramp inversely scales with weakening rate. This is an important result since it has been shown that weakening rate is correlated with the global rupture velocity of cracks [4,7]. To verify this prediction, we measured the rate of weakening from R profiles $\dot{R}_w = \frac{R_{max} - R_{min}}{\tau_w}$. As we have shown in Fig.6b, our measurements qualitatively confirms the aforementioned prediction.

We have shown, for the first time via laboratory data, the microcrack evolution over the duration of only a few microseconds derived purely from multiple AEs event data. We mapped acoustic excitations from crackling events to complex networks and then further to Ising-like chains. Using these novel tools, we elucidated the transition to fast-weakening, and we showed that the nature of this transition is governed by a defect mediated transition process. In particular, very close to the critical point of failure, the correlation length is frozen, and the density of nucleated defects scales with local stress ramp rate, implying that the Kibble-Zurek mechanism can be employed in studying acoustic crackling noises. To determine the KZM exponent in the estimation of defect density, we made extensive measurements over many reordered crackling noises from our experiments. In addition, we found that dynamic stress-time profiles involve a double-KZM theory where order-disorder-order transitions occur. With this interpretation, we could estimate the linear rate of the weakening phase, a parameter closely to global rupture velocity. While all our focus on this study was to characterize events with the definite continuous transition of S-W, nevertheless, we could recognize events with an abrupt change in the order parameter that is characteristic of the first order transitions (Fig.S18). Further study is needed to explore the frequency of these first-order "laboratory earthquakes" and the boundary or environmental conditions which favor dominant first-order crackling noises. Having multiple-acoustic cracking noises where each nodes sets in analogy with spins, it would be interesting to study the effect of flipping nodes and the concept of "spin-waves" (i.e., crystal vibrations or phonons [19]) and their interaction in the course of the S-W transition. Furthermore, future work should be focused on studying the thermal activation mechanism of nucleated kinks prior to freeze-out time and the effects of weak-planes (cleavage) in different crystals on polarization patterns of K-strings.

## Methods

**Laboratory Procedures:** We use four sets of recorded acoustic emissions (labeled as Lab.EQ1, 2, 3 and 4) from Westerly granite and Basalt rock samples (most of the analyzed events are precursor rupture fronts). The Lab.EQ1and 2 are the recorded multi-stationary acoustic waveforms from evolution of frictional rock-interfaces of Westerly Granite samples. The interfaces were in dry conditions with smooth (saw-cut) and naturally rough surfaces, respectively [20]. The evaluated events are from different stages and position and are not limited to particular stage of the tests. The Lab.EQ3 is the fast-loading experiment on a cylindrical sample of Westerly Granite (~$10^{-5}$ s$^{-1}$) at 50MPa confining pressure (approximately 2km), which is about an order faster than Lab.EQ1 and 2 [21]. Lab.EQ4 are events from Basalt samples; described in [22]. The global loading rate was $10^{-6}$ s$^{-1}$. In all of the above experiments, we reordered amplified events using 16 to 18 sensor networks in both short (discrete events) and long timescale recorders (AE records). The resolution of each recorded interval during the life-time of a waveform was ~20-50ns.

**Networks of Acoustic emission waveforms:** The concept behind studying each single acoustic excitation event is to determine the onset of the critical phase where the defects precipitate out in a matrix of elastic material. The excitation of defects in the vicinity of a moving crack tip (such as dislocations) must be reflected in each recorded acoustic event. To study each acoustic event, we use functional network theory to analyze multiple recorded waveforms. The full details of the employed algorithm can be found in [4,6-7; Section 1 of SI]. In summary, the algorithm uses a thresholded-closeness metric to establish *functional networks* where the nodes are the sensors and per each time step, a network is assigned. Per each network a set of properties - nodes' degree, modularity index and centrality measures - are extracted (see the definition in SI-section 1). The temporal evolution of Q-values in the monitored time interval then forms the Q-curves. Q(t) carries generic universal time scales, distinguishing details of micro-cracks over microsecond timescale. Following [4,12], we found three generic classes, corresponding to the following phases in Q-profiles (Fig.S.1): (1) S-phase: nucleation and main deformation phase with the signature of initial strengthening; (2) W-phase or fast-slip and; (3) D-phase or slow slip stage. To distinguish the role of the S- phase, we define the parameter such as: $R \equiv (Q_{norm.})^{-1}$ where $Q_{norm.} = \frac{Q}{Q_0}$ ($Q_0$ is the initial or rest value of Q(t)). We also use a parameter indicating the mean of local energy flow for a given network: betweenness centrality (B.C). It characterizes the importance of a node using the number of shortest paths from all nodes that pass through that node [23].

**Kibble-Zurek Mechanism:** The idea behind the KZM is to compare the relaxation time (or healing time of the system in equilibrium) with the timescale of change of the control parameter ( ). Assuming a linear change of control parameter in vicinity of the critical point $(t) = t/\tau_s$, where $\tau_s$ is the ramp time in S-phase. The relaxation or healing time we consider is an equilibrium (static or quasi-static) condition: $\tau(v) = \frac{\tau_0}{|v|^{vz}}$ and $\epsilon_z = \sim$. This determines the reaction time of the order parameter. Here, and $z$ are spatial and dynamical critical exponents, and $\tau_0$ is a characteristic timescale. The system can adiabatically follow the change imposed by the local stress ramp if relaxation time characterized by ( ) is outside the interval $\hat{t} = (\tau_0 \tau_s^{\epsilon})^{\frac{1}{1+\epsilon z}}$ centered around the transition point ($\hat{t}$ is freeze-out time).

The system will cease to maintain with the imposed change at time $\hat{t}$ before reaching the critical point and correlation length of the system is effectively frozen. The correlation length is given by [9-10]: $\hat{\xi} = \xi_0 (\tau_s/\tau_0)^{\frac{\nu}{1+\nu z}}$. Topological defects are formed with the density of one defect fragment per domain. An estimate for the resulting density of topological defects is given by [9-10]: $n \sim \frac{1}{\xi_0^2}(\tau_0/\tau_s)^{\frac{2\nu}{1+\nu z}}$. In the frozen phase, one can define an effective control parameter $\hat{v} = \hat{R} = \frac{\hat{t}}{\tau_s} = (\frac{\tau_0}{\tau_s})^{1/1+\nu}$. From plotting events in $\hat{R} - \dot{R}_c$ plane which hold $\hat{R} \sim (\dot{R}_c)^{1/1+\nu}$, we obtain $\nu z = \nu$. Then, we can estimate $\nu$ from $\hat{\xi} \sim (\dot{R}_c)^{\frac{-\nu}{1+\nu z}}$ where we measure the frozen correlation length for the given event with the rate of transition $\dot{R}_c$. This procedure leads to estimate $\nu \cong 0.5$ and $z \cong 1$ which does match with mean-field 's approximation of the scaling coefficients of the Ising-model ($Z_2$ symmetry breaking [18]). To test the KZ's scaling exponent, we analyzed events with definitive continuous transition of order parameter (Fig.S15-17) while we measure the frozen correlation length in the Ising-like chains of functional acoustic networks (Fig.S.14).

**Supplementary Information**

Supplementary information accompanies this paper.

**Acknowledgements**

We would like to acknowledge and thank B.D. Thompson (Mine Design Engineering, Kingston, Canada), for providing part of the employed data set in this work. The first author would like to acknowledge *A.del Campo* and *W.Zurek* for their encouragements and points on the results.

**Author Contributions**

All authors contributed to the analysis and writing of the manuscript.

**Competing interests statement**

The authors declare no competing financial interests.

# Figures

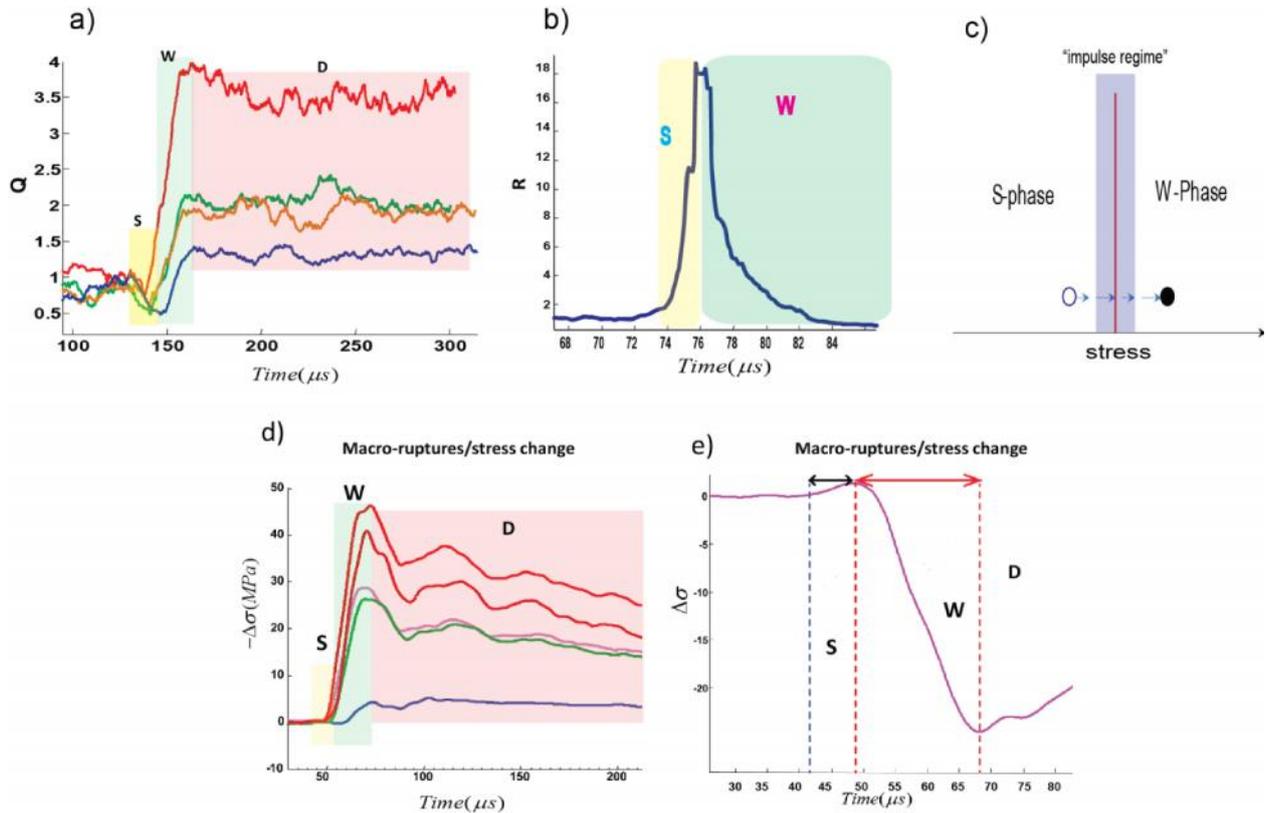

**Figure 1| Q-profiles representing dynamic crackling noises.** (**a**) Three main stages of typical acoustic crackling noises as shown in normalized Q-profiles: S-W-D phases correspond with strengthening, weakening and decelerating stages, respectively. (**b**) A typical -R profile as generated from dataset Lab.EQ1. (**c**) A schematic representation of impulse zone in transition from S to W phase. (**d**) five recorded stick-slip events with (dynamic) strain gauges measurements in centimeter scales rock-interfaces [7,21].(**e**) An event from (d) where stresses dynamically drop about 25 MPa during a stick-slip experiment.

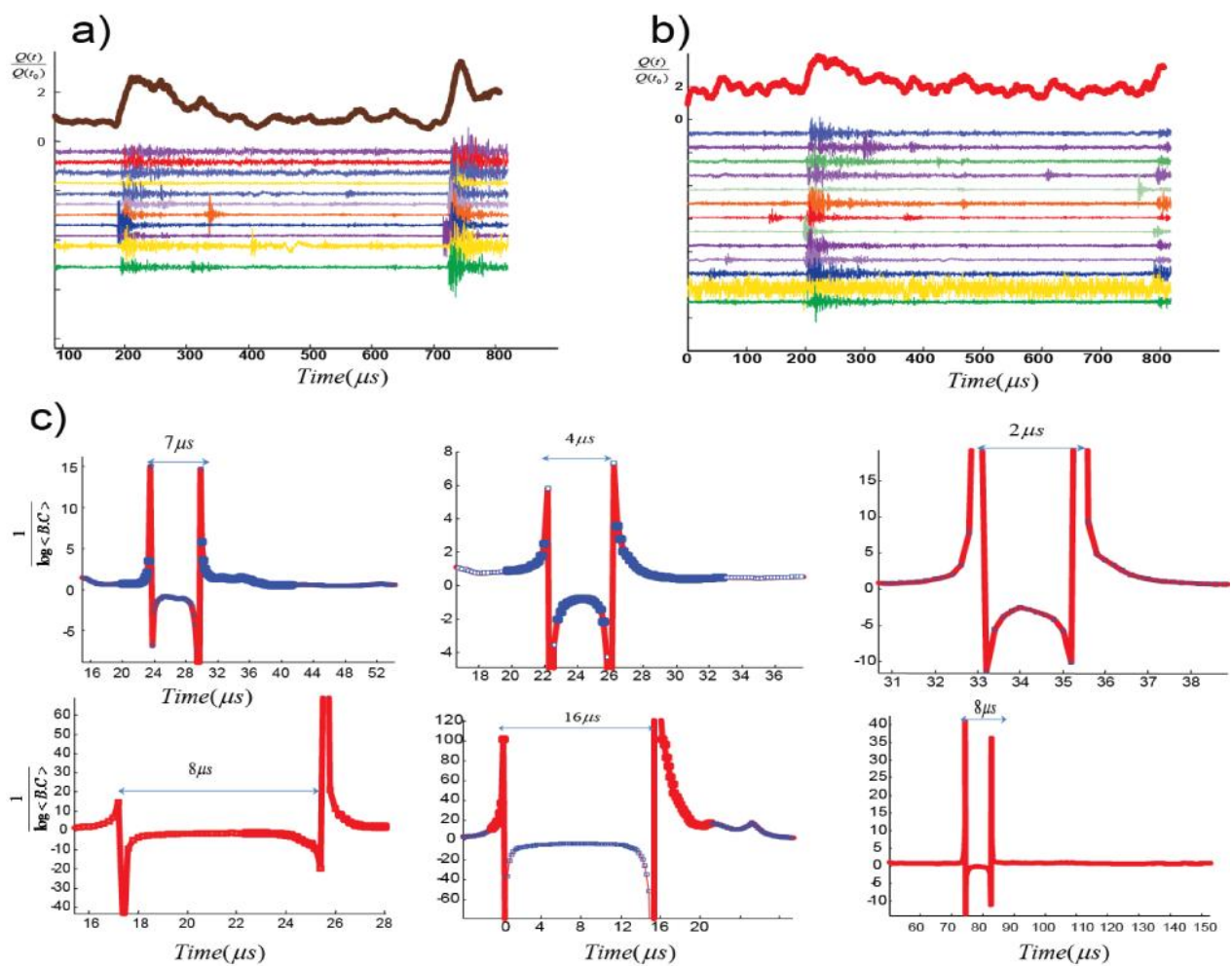

**Figure 2. (a,b) two typical acoustic emission events from cracking Granite samples (from Lab.EQ3).** We have shown scaled recorded acoustic waveforms in ~800μs and the corresponding normalized Q(t). c) Different events from our tests with the signature of inverse of mean betweeness centrality which shows divergence of the parameter in vicinity of the nucleation zone. Based on the resolution of our measurements, the total precipitate out time varies between 0.5-16μs
11

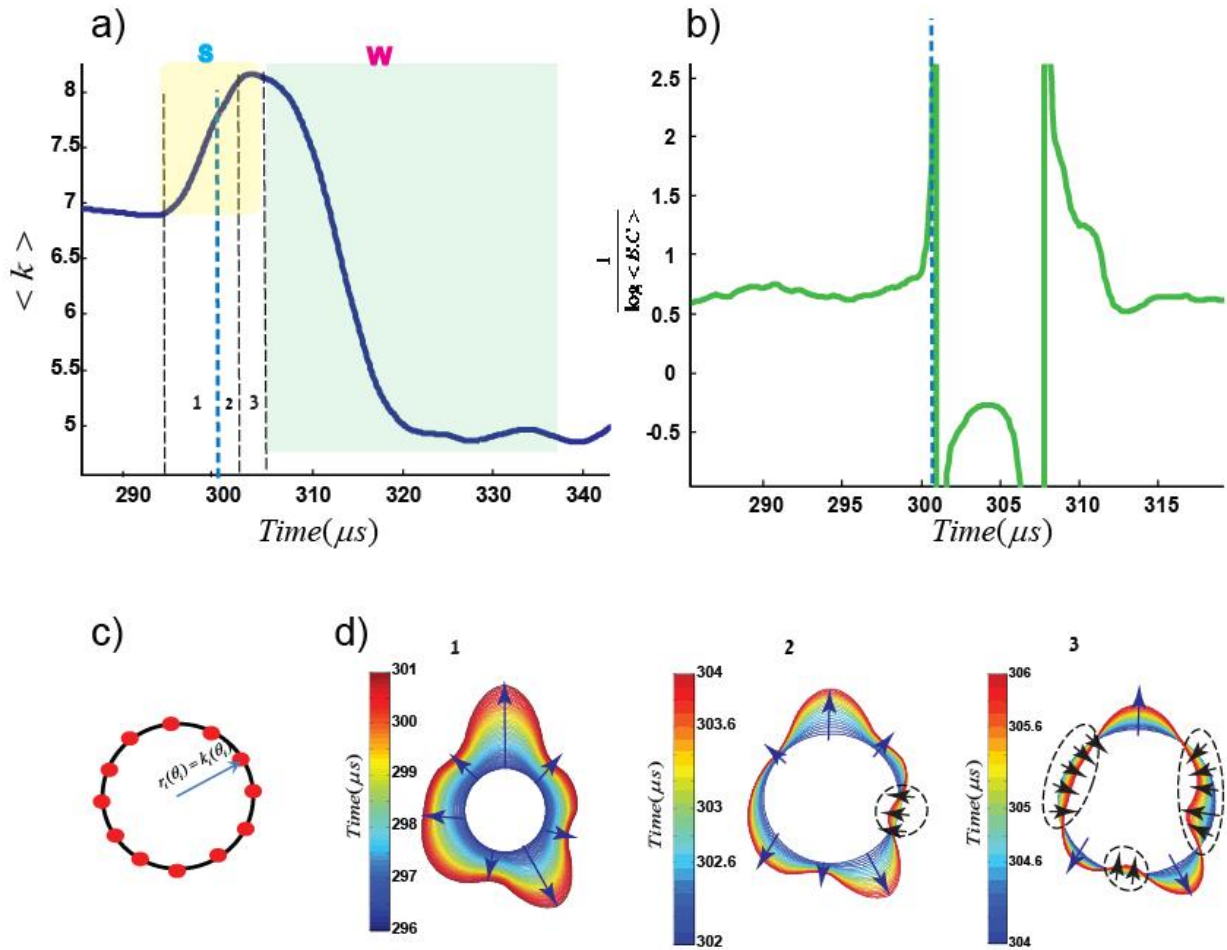

**Figure 3| Nucleation of kinks and formation of domains .** a) The mean number of edges (<k>) versus time for event#24 from Lab.EQ4. b) Transition to nucleation zone is imprinted in diverging the inverse of mean betweeness centrality (B.C). c) Schematic representation of sensors location (red filled circles) where the radius of the ring is proportional with node's degree (d) We have shown accumulated 2D spatio-temporal patterns of nodes' degree in the polar system for each time interval as in panel (a). Transition from the linear stage (1) to the non-linear regime (2+3) is indicated by the onset of local defects (black arrows), inducing formation of "*domains*". In the example reported, there are three defect-zones. The arrows are normal to strings and crumpled strings destroy long-range order in string normal. We have shown more examples in Fig.S7-S.8.

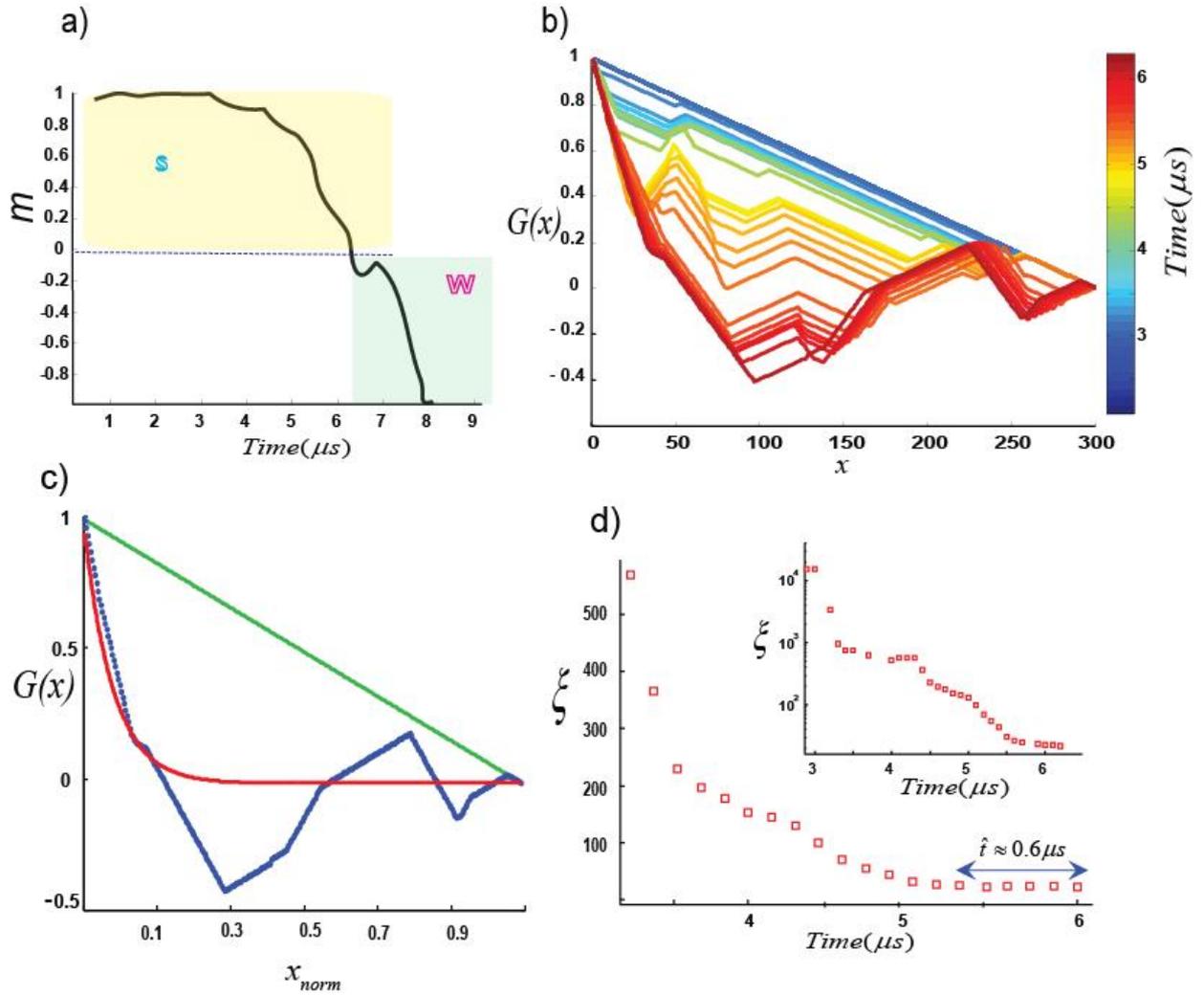

**Figure 4.** . **Continuous Phase Transition from S to W.** Results from a cracking noise #255 from Lab.EQ4 with mapping on the Ising-chain **(a)** The real–time order parameter *m* versus time. *m* goes to zero when it approaches the critical point $\}_c$ as the failure point. **(b)** Real-time cross-correlation function **.**Very close to m=0 in red, we can see nearly overlapping patterns of G(x) indicating a frozen zone of correlation length **(c)** cross-correlation function versus the normalized distance. For fully ordered state a triangular function (fully coherent) is given by green line. Approaching m=0, we can fit a correlation function (red line), to determine correlation length. For this event, the frozen correlation length is $\hat{\varsigma} \approx 0.086$. Blue points are the experimental measures of the correlation function. Before normalization, we had L=300 nodes and $\hat{\varsigma} \approx 21$ nodes. Approaching the point *m=0*, the correlation length becomes frozen as shown in **(d).** For this event , the correlation length is roughly constant for ~0.6μs. See supplementary figures S13-S15 for more examples.

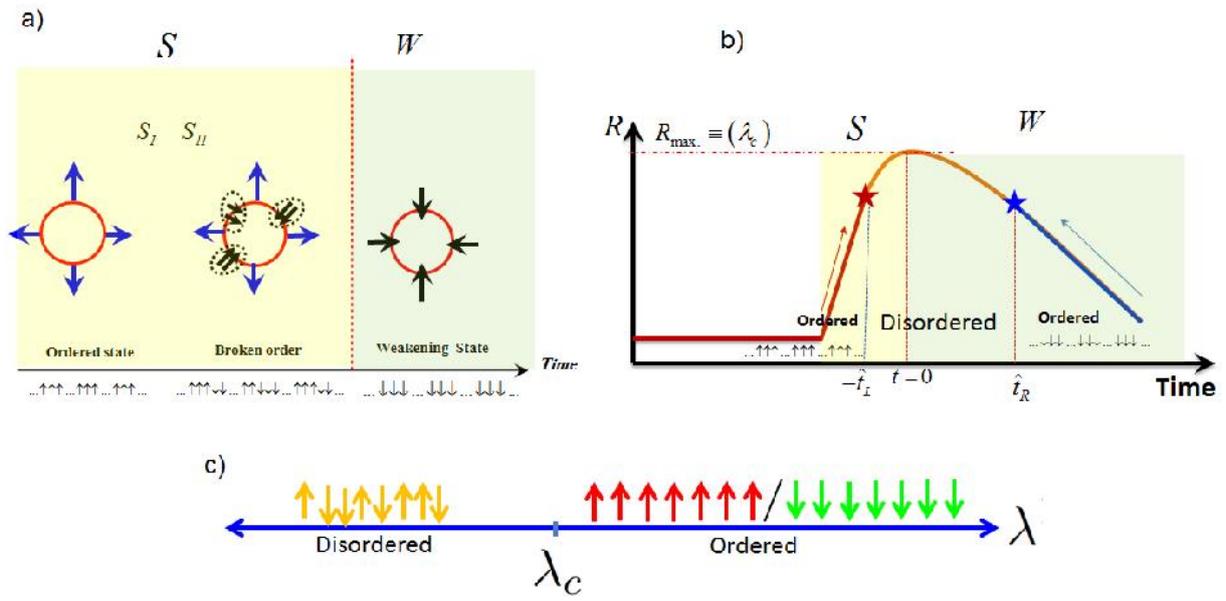

**Figure 5. | Description of Structural Phase Transition in Acoustic Excitations.** (a) Formation of domains occurs in $S_{II}$-stage while "failure" (i.e., fast-weakening regime) is characterized by undulation along the K-string. (b) Schematic of order-disorder-order transition in R-profiles: approaching from Left (S-phase) or Right (W-phase ; time reversal) to the critical point $\}_c$ results slightly different characteristics of defects. In particular, the rate of the S-ramp (linear strengthening rate; approaching from left) is higher than the rate of W-ramp (i.e. linear weakening rate; approaching from right). According to KZM, slower ramp rate marks longer freeze-out time and smaller defect density, so time-reversal approaching to the critical point $\}_c$ yields longer KZM freeze-out time : $\hat{t}_R \geq \hat{t}_L$.

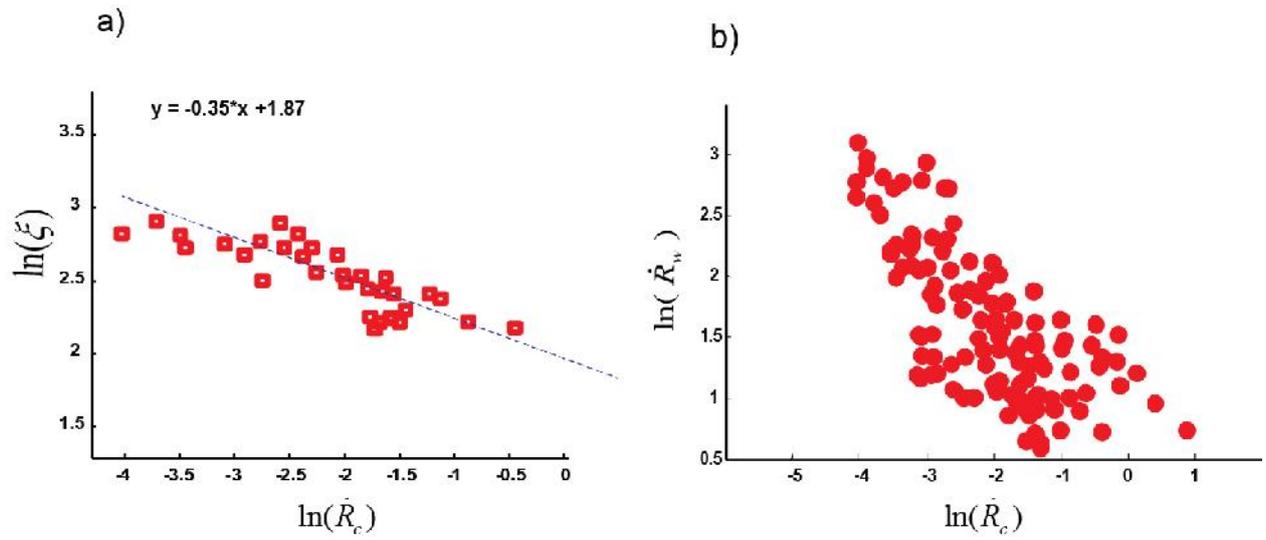

**Figure 6| Dependency of frozen correlation length (kink density) on ramping rate.** (a) Events with faster transition to their critical points induce higher defect density (i.e., shorter correlation length). Here we show typical rupture fronts from Lab.EQ3 and the size of the network is 300 nodes. We obtained $b \approx 0.35$ in $\hat{\xi} \sim (\dot{R}_c)^{-b}$ (dashed line), in agreement with the mean-field model prediction (also see Fig.S14-S16) (b) effect of local stress-ramp rates on the fast-weakening regime: Events with faster weakening rate scale with slower ramp rate. This is illustrated here via a *log-log* plot of the normalized rate of weakening $\dot{R}_w$ observed as a function of ramp-rate (events from Lab.EQ4).